\documentclass[fleqn,10pt]{wlscirep}
\usepackage[utf8]{inputenc}
\usepackage[T1]{fontenc}
\title{Machine learning assisted quantum state estimation}

\author[1]{Sanjaya Lohani}
\author[2,+]{Brian T. Kirby}
\author[2]{Michael Brodsky}
\author[1]{Onur Danaci}
\author[1,*]{Ryan T. Glasser}
\affil[1]{Tulane University, New Orleans, LA 70118, USA}
\affil[2]{United States Army Research Laboratory, Adelphi, Maryland 20783, USA}

\affil[+]{brian.t.kirby4.civ@mail.mil}
\affil[*]{rglasser@tulane.edu}

\begin{abstract}

We build a general quantum state tomography framework that makes use of machine learning techniques to reconstruct quantum states from a given set of coincidence measurements. For a wide range of pure and mixed input states we demonstrate via simulations that our method produces functionally equivalent reconstructed states to that of traditional methods with the added benefit that expensive computations are front-loaded with our system. Further, by training our system with measurement results that include simulated noise sources we are able to demonstrate a significantly enhanced average fidelity when compared to typical reconstruction methods. These enhancements in average fidelity are also shown to persist when we consider state reconstruction from partial tomography data where several measurements are missing. We anticipate that the present results combining the fields of machine intelligence and quantum state estimation will greatly improve and speed up tomography-based quantum experiments.

\end{abstract}
\begin{document}

\flushbottom
\maketitle
\thispagestyle{empty}

\section*{Introduction}
Quantum information science (QIS) is a rapidly developing field that aims to exploit quantum properties, such as quantum interference and quantum entanglement\cite{horodecki_quantum_2009}, to perform functions related to computing\cite{nielsen2015neural}, communication\cite{mattle_dense_1996}, and simulation\cite{georgescu_quantum_2014}.  
Interest in QIS has grown rapidly since it was discovered that many tasks can be performed using QIS systems either more quickly than, or which are completely unavailable to, their classical counterparts.  
In general, all QIS tasks require the support of classical computation and communication in order to coordinate, control, and interpret experimental outcomes.  
While the classical overhead needed to effectively operate and understand quantum systems is often negligible in current experimental settings, the exponential growth of parameters describing a quantum system with qubit number will quickly put substantial demands on available computing resources.  

\begin{figure}[b!] 
\centering\includegraphics[width=\linewidth]{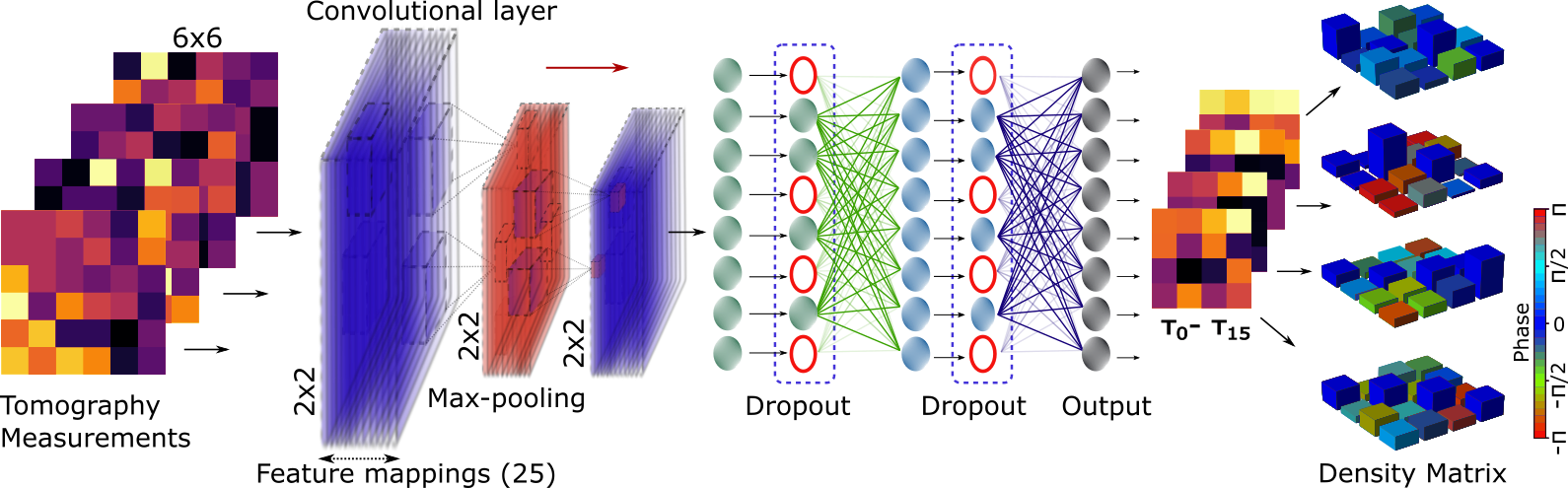}
\caption{Schematic of the robust tomography scheme with machine learning. The noisy tomography measurements are fed to the convolutional neural network, which makes predictions of intermediate $\tau$-matrices as the outputs. At the end, the predicted matrices are inverted to reconstruct the pure density matrices for the given noisy measurements.} 
\label{fig:Figure_1}
\end{figure}

Using machine learning (ML) to reduce the burden of classical information processing for QIS tasks has recently become an area of intense interest.  Examples of where this intersection is being investigated include the representation and classification of many-body quantum states\cite{carleo_constructing_2018}, the verification of quantum devices\cite{lennon2019efficiently}, quantum error correction\cite{nautrup2019optimizing}, quantum control\cite{kalantre_machine_2019}, and quantum state tomography (QST)\cite{altepeter2005photonic, zimmermann_high-resolution_2011}.  
Here we focus on QST, where a large number of joint measurements on an ensemble of identical, but completely unknown, quantum systems are combined to estimate the unknown state.  For a quantum state of dimension $d$ there are $d^2-1$ real parameters in the density matrix describing that state, and hence the resources required to measure and process the data required for QST grows quickly for systems with large dimension, such as those needed to demonstrate quantum supremacy\cite{boixo_characterizing_2018}.  Current methods for full state reconstruction from tomographic measurements scale as $O(d^4)$ even when making the simplifying assumption that all noise is Gaussian \cite{smolin_efficient_2012,qi_quantum_2013,hou_full_2016}.  
As an example how demanding this scaling is in modern experiments, the reconstruction of an 8-qubit state in \cite{haffner2005scalable} took weeks of computation time, in fact, more time than was required for data collection itself \cite{gross2010quantum}.
Recently, various deep learning approaches have been proposed for efficient state reconstruction\cite{torlai_neural-network_2018,carrasquilla_reconstructing_2019,torlai_latent_2018,xin_local-measurement-based_2018,palmieri_experimental_2019,lohani2019dispersion} with some techniques indicating a scaling of $O(d^3)$\cite{xu_neural_2018}. 

In this paper we implement a convolutional neural network (CNN) to reduce the computational overhead required to perform full QST.
Our system is shown via simulated measurements to construct equivalent density matrices to traditional methods of state estimation to a high degree of accuracy.
Our QST system has the distinct benefit that all significant computations can be performed ahead of time on a standalone computer with the final result deployed on more modest hardware.
Further, in the setting where tomographic measurements are noisy or incomplete, we are able to demonstrate a significant enhancement in average fidelity over typical reconstruction methods by training our QST system with simulated noise ahead of time.
These results constitute a significant step toward the implementation of high-speed QST systems for applications requiring high-dimensional quantum systems.

The design of our QST setup is shown schematically in Fig. \ref{fig:Figure_1}. A series of noisy and potentially incomplete measurements performed on a given density matrix are simulated, which are then fed to the input layer of a CNN. Examples of the noisy tomography are shown as the tomography measurements in Fig. \ref{fig:Figure_1} (left-side). Then the CNN makes the prediction of $\tau$-matrices (which are discussed in the following section) as the output. Finally, the output is inverted, resulting in a valid density matrix. Examples of the reconstructed density matrices are shown in Fig. \ref{fig:Figure_1} (right-side). This process is repeated many times for various sizes of random measurements, strengths of noise, missing measurements.
The average fidelity ($F$) of the setup is calculated and compared to the fidelity when a non-machine learning method is used.

\section*{Results}
The general setup of our CNN is depicted in Fig. \ref{fig:Figure_1}, which consists of feature mappings, max pooling, and dropout layers\cite{rawat2017deep}.
More specifically, the two dimensional convolutional layer has a kernel of size of $2\times2$, stride length of 1, 25 feature mappings, zero padding, and a rectified linear unit (ReLU) activation function. 
The max-pooling layer is two-dimensional and has a kernel of size $2\times2$ with stride length of 2 which halves the dimension of the inputs, which is further followed by a convolutional layer with the same parameters as discussed above. Next, we attach a fully connected layer (FCL) with 720 neurons, and the ReLU activation. Then we have a dropout layer with a rate of 50$\%$, which is followed by another FCL with 450 neurons, and the ReLU activation. Similarly, after this we attach, again, a dropout layer with a rate of 50$\%$, which is finally connected with an output layer with 16 neurons. Note that the hyperparameters of the CNN are manually optimized as discussed in\cite{lohani_use_2018}.
Furthermore, the network is designed such that the output (firing of 16 neurons) comprises the elements of the $\tau$-matrix (see Method), which can be listed as [$\tau_0$,$\tau_1$,$\tau_2$,$\tau_3$, .. .. .., $\tau_{15}$]. Next, the list of 16 elements is re-arranged to form a lower triangular matrix as given in equation \ref{eqn:lower_t}
\begin{equation}
    \tau_{predicted} = [\tau_0,\tau_1,\tau_2,\tau_3, .. .. .., \tau_{15}] \rightarrow \begin{bmatrix}
                                       \tau_0 & 0& 0& 0\\
                                       \tau_4+i\tau_5 & \tau_1 &0 &0\\
                                       \tau_{10}+i\tau_{11} & \tau_6+i\tau_7 &\tau_2 &0\\
                                       \tau_{14}+i\tau_{15} & \tau_{12}+i\tau_{13} &\tau_8+i\tau_9 &\tau_3\\
    \end{bmatrix},
    \label{eqn:lower_t}
\end{equation}
which is, finally, compared with the target ($\tau_{target}$) for the given measurements (see Method) in order to find the mean square loss. 
We optimize the loss using adagrad-optimizer (learning rate of 0.008) of tensorflow\cite{martin_abadi_tensorflow:_2015}. Additionally, at the end of an epoch (one cycle through the entire training set), the network makes the  $\tau$-matrix prediction for the unknown (test) noisy measurements, which is later inverted to give the tomography and fidelity of the setup as given by equation \ref{eqn:tomo},
\begin{equation}
    \rho_{pred} = \frac{\tau_{pred}^\dagger\tau_{pred}}{Tr(\tau_{pred}^\dagger\tau_{pred})}, \qquad F = \Big|Tr\sqrt{\sqrt{\rho_{pred}}\rho_{targ}\sqrt{\rho_{pred}}}\Big|^2
    \label{eqn:tomo}
\end{equation}
where $\rho_{pred}$ and $\rho_{targ}$ represent the predicted and target density matrices, respectively. 
The form of equation \ref{eqn:tomo} guarantees that the network always makes predictions which are physically valid \cite{james_measurement_2001}.
Note that the conversion of $\tau$-matrices to their corresponding density matrices and evaluation of the fidelity are inbuilt to the network architecture, so there is no separate post-processing unit.
\begin{figure}[h!] 
\centering\includegraphics[width=\linewidth]{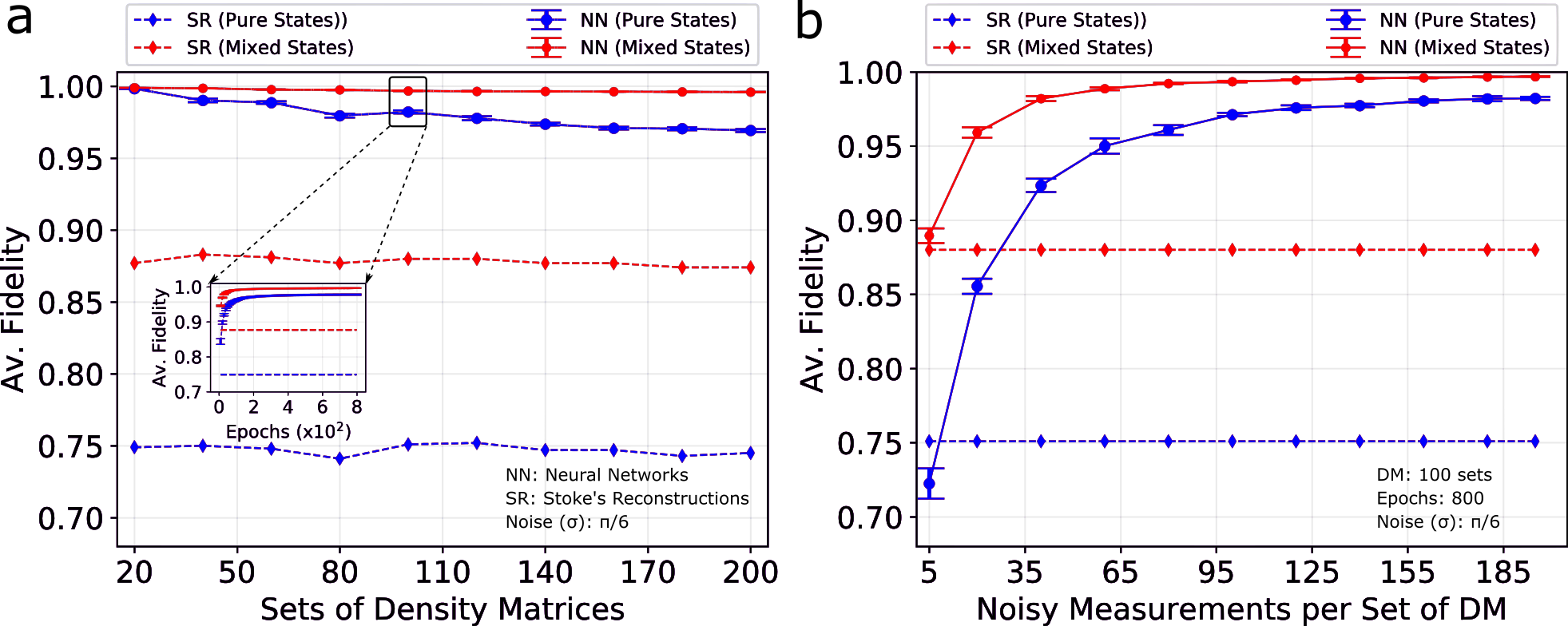}
\caption{(a) Average fidelity of the reconstructed density matrices (DM) for the unknown noisy measurements versus number of density matrices used to train the networks. Similarly, the progressive average fidelity versus number of epochs with 100 sets of density matrices is shown in the inset. (b) Average fidelity versus number of noisy measurements per target density matrix. The error bars represent the one standard deviation from the mean value.}
\label{fig:Figure_2}
\end{figure}

First we evaluate the average fidelity with respect to number of sets of density matrices used in the network for both pure and mixed states. In order to generate training and test sets, we randomly create 200 density matrices and their corresponding $\tau$-matrices (see ``Method''), again for both pure and mixed states. After this we randomly simulate the 200 noisy ($\sigma\,=\,\pi/6$) tomography measurements (each measurement contains 36 projections as described in ``Method'') for each of the $\tau$-matrices, for a total of 40,000 sets (see ``Method''). 
We then split each set of 200 noisy measurement results per $\tau$-matrix into training and test sets (unknown to the network) with sizes of 195 and 5, respectively.  For example, if we are working with 80 random density matrices ($\tau$-matrices) then 195 out of 200 noisy tomography measurement data sets per density matrices, i.e, a total of 15,600 $(80\times195)$, are used to train the network and a total of 400 ($80\times5$), are used to test the network. Note that in order to efficiently train the networks, we implement the batch optimization technique with a batch size of 4 for all the calculations discussed in the paper. With these training sets and hyper-parameters the CNN is then pre-trained up to 800 epochs. 

For comparison with standard techniques, we also implement the Stokes reconstruction method\cite{james_measurement_2001} (see ``Method'').  
The average fidelity is found to be significantly enhanced when the CNN is used (solid line) over the Stokes technique (dotted line) for the various number of sets of density matrices is shown in Fig. \ref{fig:Figure_2} (a). Note that we run the same training and testing process 10 times with different (random) initial points, in order to gather statistics (shown by the error bars). In the case of 20 sets of density matrices, we find a remarkable improvement in average fidelity from 0.749 to 0.998 with a standard deviation of $2.9\times10^{-4}$, and 0.877 to 0.999 with a standard deviation of $1.21\times10^{-4}$ for the pure states (blue curves) and mixed states (red curves), respectively. Similarly, even for the larger sets of 200 density matrices we find an enhancement of 
0.745 to 0.969 with a standard deviation of $1.03\times10^{-3}$, and 0.874 to 0.996 with a standard deviation of $2.07\times10^{-4}$ for the pure states and mixed states, respectively.
These results not only demonstrate an improved  fidelity when compared to Stokes reconstruction but also approach the theoretical maximum value of unity.
Additionally, improvement in average fidelity of the generated density matrices for unknown noisy tomography measurements per each training epoch is shown in the inset of Fig.  \ref{fig:Figure_2} (a). The average fidelity is found to be saturated after 500 epochs.

We have also investigated how the number of noisy training sets per random density matrix impacts the effectiveness of our system.  To do this we fix the number of sets of density matrices at 100 and vary the number of noisy measurements per set (in the previous paragraph, and Fig. \ref{fig:Figure_2} (a), this was fixed at 195). For testing purposes we use the same 5 noisy measurement sets per random density matrix which were used to create Fig. \ref{fig:Figure_2} (a). As expected, the average fidelity improves noticeably as the number of noisy measurement training sets per random density matrix is increased, as shown in Fig. \ref{fig:Figure_2} (b). 
Specifically, the average fidelity improves from 0.751 to 0.982 with a standard deviation of $1.04\times10^{-3}$, and 0.88 to 0.996 with a standard deviation of $2.1\times10^{-4}$ for the pure states and mixed states, respectively. Additionally, even when we only train on simulated noise 40 times per random density matrix the average fidelity still increases from 0.751 to 0.923, and 0.88 to 0.982 with a standard deviation of $4.5\times10^{-3}$ and $1.6\times10^{-3}$, respectively, for the pure and mixed states.  
\begin{figure}[h!] 
\centering\includegraphics[width=\linewidth]{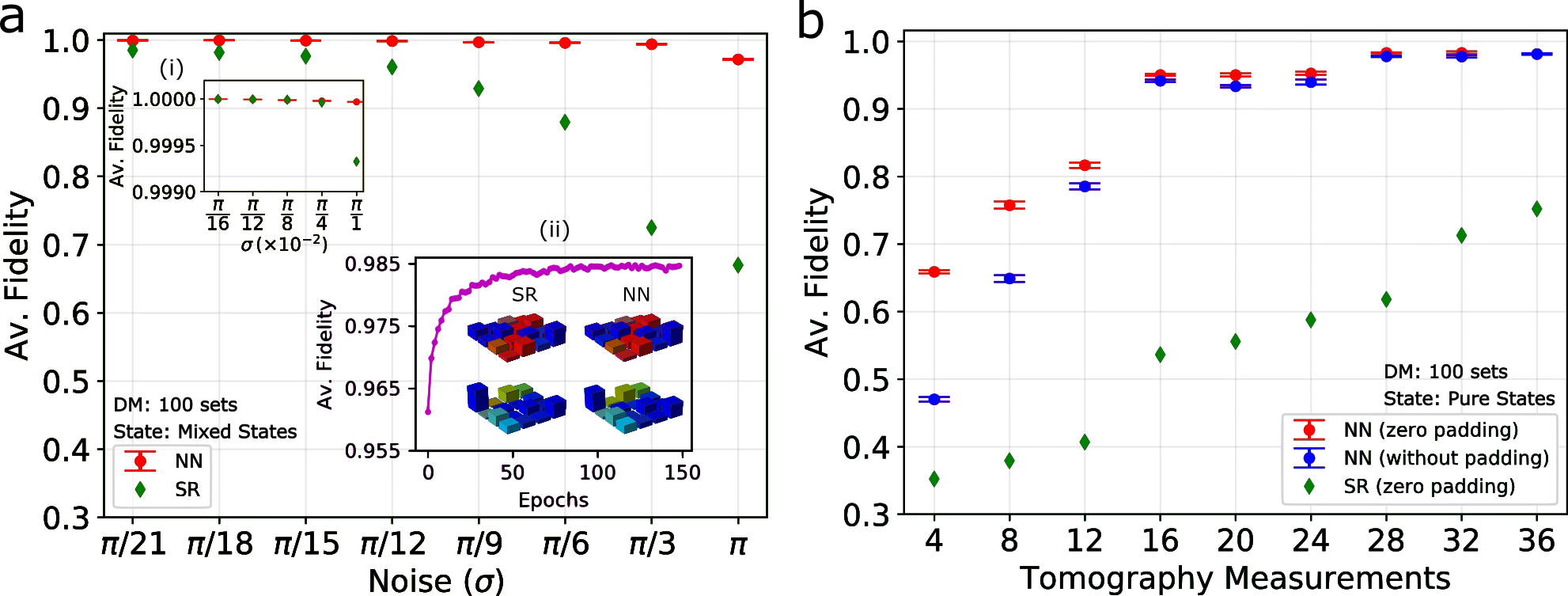}
\caption{(a) Average fidelity versus the amount of noise present in the tomography measurements. Here $\sigma$ represents the strength of the noise. Similarly, (a-i) the average fidelity for the less noisy sets are shown in the inset, (a-ii) the average fidelity versus epochs for completely unknown noiseless tomography measurements. SR (left-column): quantum states generated with Stokes reconstruction method, and NN (right-column): quantum states generated with CNN. (b) Average fidelity versus size of tomography measurements (number of projection operators) out of the total of 36 complete tomography set. In the both cases, the error bars represent the one standard deviation from the mean value.}  
\label{fig:Figure_3}
\end{figure}

In order to investigate the robustness of our system, we now vary the strength ($\sigma$) of noise used to both train and test our CNN.  Specifically, we vary the noise strength from strong, $\sigma\,=\,\pi$, to weak, $\sigma\,=\,\pi/21$. For each $\sigma$ value, we fix the number of sets of density matrix at 100 and randomly generate 200 noisy tomography measurements per set of density matrices resulting in a total of 20,000. As previously discussed, 195 (total of 19,500) and 5 (total of 500) out of the 200 per set of density matrices for the given noise are used as the training and test set, respectively. Note that we separately train the CNN for each different value of the noise. With the CNN pre-trained up to 500 epochs, a significant improvement in average fidelity of the generated density matrices with the CNN (red dots) over the conventional method (green dots) at various strengths of noise is shown in Fig. \ref{fig:Figure_3}(a). We find a significant enhancement in average fidelity from 0.669 to 0.972 with a standard deviation of $7.8\times10^{-4}$, and 0.985 to 0.999 with a standard deviation of $4.96\times10^{-5}$ for the strong noise strength of $\sigma\,=\,\pi$, and the weak noise strength of $\sigma\,=\,\pi/21$, respectively. 
Similarly, for weaker strengths of noise, we show the average fidelity of the generated quantum states with the CNN begins to converge with the conventional method as shown in the inset of Fig. \ref{fig:Figure_3} (a-i).  We find the average fidelity from the CNN generation method as well as the conventional method for the noise strengths of $\pi/800$, $\pi/1200$, and $\pi/1600$ converge to unity.  This can be considered qualitative evidence that our CNN approach to quantum state reconstruction is effectively equivalent to Stokes reconstruction in the absence of measurement noise. In order to further illustrate the efficacy of the CNN, we simulate 60,000 random tomography data sets without measurement noise. Note that the simulated 60,000 tomography measurements are random and unique. As before, the total set is divided into a training set with 55,000 measurements, and a testing set with 5,000 measurements. The tomography measurements in the testing set are completely unknown to the network. The average fidelity of the generated quantum states via the CNN per epoch for the unknown measurement data is shown in  the inset of Fig. \ref{fig:Figure_3} (a-ii). We find the generated quantum states from the CNN (NN: right-column) for the blind test data are functionally equivalent to Stokes reconstruction (SR: left-column) as shown in the inset of Fig. \ref{fig:Figure_3} (a-ii).

Lastly, we investigate how our CNN can handle the experimental scenario where some fraction of the 36 total tomography measurements is missing.  
Since the remaining bases measurements are not guaranteed to span the total 2-qubit Hilbert space, there is a priori reason to assume our CNN should have an advantage over 
Stokes reconstruction for this problem. 
For this analysis we use data with 100 sets of density matrices, a noise strength of $\sigma\,=\,\pi/6$, and the same training and testing data structure as previously discussed.  
However, in order to simulate missing measurement points we reduce the number of features in the input data.  
For example, in the extreme case of only using four projective measurements the input consists of only 4 feature float points over the $6\times6$ available space. The remaining 32 spaces are filled with 0 (zero padding). Similarly, for 8 projectors, 28 places are filled with 0; for 12 projectors, 24 places are filled with 0, and so on.
For the sake of comparison we also perform zero-padding on the matrices for use with Stokes reconstruction.  
With training up to 500 epochs, we find an improvement in the average fidelity of the generated density matrices with the CNN (red dots) over the conventional Stokes technique (green dots) for every available size of the tomography measurements (projectors) as shown in Fig. \ref{fig:Figure_3}(b). Note that the error bars represent one standard deviation away from the mean value. We find a significant enhancement in the average fidelity from 0.61 to 0.9827 with a standard deviation of $1.08\times10^{-3}$; from 0.532 to 0.95 with a standard deviation of $1.5\times10^{-3}$, and from 0.352 to 0.658 with a standard deviation of $2.3\times10^{-3}$ for the measurement size of 28, 16, and 4, respectively. In addition, we find an enhancement in the average fidelity even without zero padding in the input data with the CNN, which are shown by blue dots in Fig. \ref{fig:Figure_3} (b). 

\section*{Discussion}
We demonstrate quantum state reconstruction directly from projective measurement data via machine learning techniques.  
Our technique is qualitatively shown to reproduce the results of standard reconstruction methods when ideal projective measurement results are assumed.  
Further, by specifically training our network to deal with a common source of error in projective measurement data, that of measurement basis indeterminacy, we show a significant improvement in average fidelity over that of standard techniques.  
Lastly, we also consider the common situation where some number of the projective measurements are unsuccessfully performed, requiring the reconstruction of a density matrix from partial projective data.  
This situation is particularly troublesome as the final set of projectors used to collect data are unlikely to span the full Hilbert space.
For this scenario we find a dramatic improvement in the average reconstruction fidelity even when only 4 of the total 36 measurements are considered.
These results clearly demonstrate the advantages of using neural networks to create robust and portable QST systems.

\section*{Methods}
\subsection*{Generating pure states}
We define the horizontal and vertical polarization states as $H$ and $V$, respectively, which are given by equation \ref{eqn:HV},
\begin{equation}
   |H\rangle\,=\,\begin{bmatrix}
          1\\
          0\\
         \end{bmatrix}, \quad\, \text{and}\, \quad 
         |V\rangle\,=\,\begin{bmatrix}
         0\\
         1\\
         \end{bmatrix}.
\label{eqn:HV}
\end{equation}
In order to generate the pure states, we use Haar measure to simulate $4\times4$ random unitary matrices $u$. Then we use the first column of the simulated random unitary matrices as the coefficients of the pure states as in equation \ref{eqn:psi} 

\begin{equation}
    |\psi\rangle = u_{00}|HH\rangle + u_{10}|HV\rangle + u_{20}|VH\rangle + u_{30}|VV\rangle,
    \label{eqn:psi}
\end{equation}
where $u_{ij}$ represents the $i^{th}$ row and $j^{th}$ column of the random unitary matrix ($u$), $|HH\rangle$, $|HV\rangle$, $VH\rangle$, and $|VV\rangle$ are the tensor products $|H\rangle\otimes |H\rangle$, $|H\rangle\otimes |V\rangle$, $|V\rangle\otimes |H\rangle$, and $|V\rangle\otimes |V\rangle$, respectively. Note that we add a tiny perturbation term $\epsilon\,(1\times10^{-7})$ to the simulated pure states as given in equation \ref{eqn:epsilon} to avoid the possible convergent issue under Cholesky decomposition of the pure state density matrix ($\rho_p$)\cite{higham_analysis_1990},
\begin{equation}
    \rho_{pure} = (1-\epsilon)|\psi\rangle\langle\psi| + \frac{\epsilon}{4}I.
    \label{eqn:epsilon}
\end{equation}
\subsection*{Generating mixed states}
First we simulate the random matrix from the Ginibre ensemble\cite{forrester_eigenvalue_2007} as given in equation \ref{eqn:gini},
\begin{equation}
    G = N\big(0,1,[4,4]\big)\,+\,i\,N\big(0,1,[4,4]\big)
    \label{eqn:gini}
\end{equation}
where $N\big(0,1,[4,4]\big)$ represents the random normal distribution of size of $4\times 4$ with zero mean and unity variance. Finally, the random density matrix ($\rho_m$) using the Hilbert-Schmidt metric\cite{ozawa_entanglement_2000} is given by equation \ref{eqn:mix}
\begin{equation}
    \rho_{mix} = \frac{GG^\dagger}{Tr(GG^\dagger)}.
    \label{eqn:mix}
\end{equation}
Where $Tr$ represents the trace of a matrix. 
\subsection*{Simulating tomography measurements}
Here we simulate the exact sequence of the tomography measurements used by the Nucrypt entangled photon system \cite{wang_robust_2009}. In addition to $|H\rangle$ and $|V\rangle$, we now define a diagonal ($|D\rangle$), anti-diagonal ($|A\rangle$), right circular ($|R\rangle$), and left circular ($|L\rangle$) polarization states, which are given in equation \ref{eqn:darl}
\begin{equation}
    |D\rangle = \frac{1}{\sqrt{2}}(|H\rangle+|V\rangle),\,\quad
    |A\rangle = \frac{1}{\sqrt{2}}(|H\rangle-|V\rangle),\,\quad
    |R\rangle = \frac{1}{\sqrt{2}}(|H\rangle+i\,|V\rangle),\,\quad
    |L\rangle = \frac{1}{\sqrt{2}}(|H\rangle-i\,|V\rangle).
    \label{eqn:darl}
\end{equation}
Furthermore, in order to simulate the experimental scenarios, we introduce the 36 projectors as given by equation \ref{eqn:proj} in the exact order of the Nucrypt's coincidence measurements,
\begin{equation}
    P = \begin{bmatrix}
        h\otimes h  & h\otimes v & v\otimes v & v\otimes h & v\otimes r & v\otimes l\\
        h\otimes l  & h\otimes r & h\otimes d & h\otimes a & v\otimes a & v \otimes d\\
        a\otimes d & a\otimes a & d\otimes a & d \otimes d & d\otimes r & d\otimes l\\
        a\otimes l & a\otimes r &a\otimes h &a\otimes v & d\otimes v & d\otimes h\\
        r\otimes h & r\otimes v & l\otimes v & l\otimes h & l\otimes r & l\otimes l\\
        r\otimes l & r\otimes r & r\otimes d & r\otimes a & l\otimes a & l\otimes d\\
        \end{bmatrix},
    \label{eqn:proj}
\end{equation}
where $h\,=\,|H\rangle\langle H|$, $v\,=\,|V\rangle\langle V|$, $d\,=\,|D\rangle\langle D|$, $a\,=\,|A\rangle\langle A|$, $r\,=\,|R\rangle\langle R|$, and $l\,=\,|L\rangle\langle L|$. Therefore, the perfect tomography measurements (without any noise or rotations), M,  given that any density matrix $\rho$ are calculated using equation \ref{eqn:m}
\begin{equation}
    M = Tr(\rho\,P[i,j]); \quad \textrm{for}\,\quad i,j\,=\, 0,\,1,\,2,\,3,\,4,\,5.
    \label{eqn:m}
\end{equation}
Next, we discuss adding noise to the measurements, $M$. In order to do this, we introduce arbitrary rotations to the operators defined in equation \ref{eqn:proj} by making use of the unitary rotational operator (U) as given in equation \ref{eqn:rot}
\begin{equation}
    U(\vartheta,\varphi,\xi) = \begin{bmatrix}
    e^{i\varphi/2}cos(\vartheta) & -i\,e^{i\xi}sin(\theta)\\
    -i\,e^{-i\xi}sin(\vartheta) & e^{-i\varphi/2}cos(\vartheta)
    \end{bmatrix}, \quad \vartheta,\varphi,\xi \in N(0,\sigma).
    \label{eqn:rot}
\end{equation}
Note that we randomly sample $\vartheta,\varphi,\xi$ from the normal distribution with zero mean and $\sigma^2$ variance. Finally, we simulate the tomography measurements under the noisy environment as given by equation \ref{eqn:proj_noise}
\begin{equation}
    P_{noise} = \begin{bmatrix}
        h\otimes UhU^\dagger  & h\otimes UvU^\dagger & v\otimes UvU^\dagger & v\otimes UhU^\dagger & v\otimes UrU^\dagger & v\otimes UlU^\dagger\\
        h\otimes UlU^\dagger  & h\otimes UrU^\dagger & h\otimes UdU^\dagger & h\otimes UaU^\dagger & v\otimes UaU^\dagger & v \otimes UdU^\dagger\\
        a\otimes UdU^\dagger & a\otimes UaU^\dagger & d\otimes UaU^\dagger & d \otimes UdU^\dagger & d\otimes UrU^\dagger & d\otimes UlU^\dagger\\
        a\otimes UlU^\dagger & a\otimes UrU^\dagger &a\otimes UhU^\dagger &a\otimes UvU^\dagger & d\otimes UvU^\dagger & d\otimes UhU^\dagger\\
        r\otimes UhU^\dagger & r\otimes UvU^\dagger & l\otimes UvU^\dagger & l\otimes UhU^\dagger & l\otimes UrU^\dagger & l\otimes UlU^\dagger\\
        r\otimes UlU^\dagger & r\otimes UrU^\dagger & r\otimes UdU^\dagger & r\otimes UaU^\dagger & l\otimes UaU^\dagger & l\otimes UdU^\dagger\\
        \end{bmatrix}.
    \label{eqn:proj_noise}
\end{equation}
\subsection*{Stokes reconstruction}
To compare our system to a non-machine learning and non-adaptive technique, we use the Stokes reconstruction method for the given set of tomography measurements $M_{6\times6}$ (pure/noisy). 
We express the Stokes reconstruction of the density matrix as 
\begin{equation}
\begin{split}
    \rho_{recons} = \frac{1}{4}(s_{00}I\otimes I+
                     s_{01}I\otimes\sigma_{x}+
                     s_{02}I\otimes\sigma_{y}+
                     s_{03}I\otimes\sigma_{z}+
                     s_{10}\sigma_{x}\otimes I+
                     s_{20}\sigma_{y}\otimes I+
                     s_{30}\sigma_{z}\otimes I+
                     s_{11}\sigma_{x}\otimes \sigma_{x}+\\
                     s_{12}\sigma_{x}\otimes \sigma_{y}+
                     s_{13}\sigma_{x}\otimes \sigma_{z}+
                     s_{21}\sigma_{y}\otimes \sigma_{x}+
                     s_{22}\sigma_{y}\otimes \sigma_{y}+
                     s_{23}\sigma_{y}\otimes \sigma_{z}+
                     s_{31}\sigma_{z}\otimes \sigma_{x}+
                     s_{32}\sigma_{z}\otimes \sigma_{y}+
                     s_{33}\sigma_{z}\otimes \sigma_{z}),
\end{split}
\label{eqn:stoke}
\end{equation}
where $\sigma_i$ for $i \in \{x,y,z\}$ are the Pauli matrices and the parameters $s_{lk}$ for $l,k\,\in \,\{0,1,2,3\}$ for the given 36 tomography measurements are given by equation \ref{eqn:stoke_para}.
\begin{equation}
\begin{split}
    s_{00} = M[0,0] + M[0,1] + M[0,3] + M[0,2]; \quad
    s_{11} = M[2,3] - M[2,2] - M[2,0] + M[2,1]; \quad\\
    s_{12} = M[2,3] - M[2,5] - M[3,1] + M[3,0];\quad
    s_{13} = M[3,5] - M[3,4] - M[3,2] + M[3,3]; \quad\\
    s_{21} = M[5,2] - M[5,3] - M[5,5] + M[5,4]; \quad
    s_{22} = M[5,1] - M[5,0] - M[4,4] + M[4,5];\quad\\
    s_{23} = M[4,0] - M[4,1] - M[4,3] + M[4,2]; \quad
    s_{31} = M[1,2] - M[1,3] - M[1,5] + M[1,4]; \quad\\
    s_{32} = M[1,1] - M[1,0] - M[0,4] + M[0,5];\quad
    s_{33} = M[0,0] - M[0,1] - M[0,3] + M[0,2]; \quad\\
    s_{01} = M[2,3] - M[2,2] + M[2,0] - M[2,1];\quad
    s_{02} = M[5,1] + M[4,4] - M[5,0] - M[4,5];\quad\\
    s_{03} = M[0,0] - M[0,1] + M[0,3] - M[0,2]; \quad
    s_{10} = M[2,3] + M[2,2] - M[2,0] - M[2,1]; \quad\\
    s_{20} = M[5,1] - M[4,4] + M[5,0] - M[4,5];\quad
    s_{30} = M[0,0] + M[0,1] - M[0,3] - M[0,2].\quad
\end{split}  
\label{eqn:stoke_para}
\end{equation}
\subsection*{Generating the $\tau$-matrix}
In order to evaluate the $\tau$-matrix for the given set of density matrices ($\rho$), we use the matrix decomposition method discussed in \cite{james_measurement_2001}, which is given by equation \ref{eqn:tau},
\begin{equation}
    \tau = \begin{bmatrix}
                 \sqrt{\frac{Det(\rho)}{m_1^{00}}} & 0 & 0 & 0\\
                 \frac{m_1^{01}}{\sqrt{m_1^{00}m_2^{00,11}}} & \sqrt{\frac{m_1^{00}}{m_2^{00,11}}}& 0 & 0\\
                 \frac{m_2^{01,12}}{\sqrt{\rho^33}\sqrt{m_2^{00,11}}}&\frac{m_2^{00,12}}{\sqrt{\rho^{33}}\sqrt{m_2^{00,11}}}&\sqrt{\frac{m_2^{00,11}}{\rho^{33}}}
                 & 0\\
                 \frac{\rho^{30}}{\sqrt{\rho^{33}}} &\frac{\rho^{31}}{\sqrt{\rho^{33}}}&\frac{\rho^{32}}{\sqrt{\rho^{33}}}&\sqrt{\rho^{33}}
                \end{bmatrix}
\label{eqn:tau}
\end{equation}
where $m_1^{ij}$ for $i,j \in \{0,1,2,3\}$, and $m_2^{pq,rs}$ $(p\neq r$ and $q\neq s)$ for $p,q,r,s \in \{0,1,2,3\}$ are the first and second minor of $\rho$, respectively.

\bibliography{density}

\section*{Acknowledgements}
Research was sponsored by the Army Research Laboratory and was accomplished under Cooperative Agreement Number W911NF-19-2-0087. The views and conclusions contained in this document are those of the authors and should not be interpreted as representing the official policies, either expressed or implied, of the Army Research Laboratory or the U.S. Government. The U.S. Government is authorized to reproduce and distribute reprints for Government purposes notwithstanding any copyright notation herein.

\section*{Author contributions statement}
S.L. developed the neural networks and ran all simulations.  R.T.G. and B.T.K. conceived of and led the project.  S.L., R.T.G., and B.T.K. wrote the manuscript. All authors contributed to the discussions and interpretations of the results.

\section*{Data availability}
The data that support the findings of this study are available from the corresponding authors on reasonable request.

\section*{Competing interests}
The authors declare no competing interests.

\end{document}